\documentclass[%
reprint,
superscriptaddress,
 amsmath,amssymb,
 aps,
]{revtex4-1}
\usepackage{graphicx}
\usepackage{dcolumn}
\usepackage{bm}
\usepackage{mhchem}
\usepackage{braket}
\usepackage{amsfonts}
\usepackage{amssymb}

\begin{document}

\title{Phonon predictions with E(3)-equivariant graph neural networks}

\author{Shiang Fang}
\altaffiliation{These authors contributed equally}
\affiliation{Department of Physics, Massachusetts Institute of Technology, Cambridge, MA 02139, USA}

\author{Mario Geiger}
\altaffiliation{These authors contributed equally}
\affiliation{Department of Electrical Engineering and Computer Science, Massachusetts Institute of Technology Cambridge, MA 02139, USA}

\author{Joseph G. Checkelsky}
\affiliation{Department of Physics, Massachusetts Institute of Technology, Cambridge, MA 02139, USA}

\author{Tess Smidt}
\email{tsmidt@mit.edu}
\affiliation{Department of Electrical Engineering and Computer Science, Massachusetts Institute of Technology Cambridge, MA 02139, USA}

\date{\today}

\def\thefootnote{*}\footnotetext{\textbf{Equal Contribution. Order is random.}\\}\def\thefootnote{\arabic{footnote}}

\maketitle

\section{Abstract}
We present an equivariant neural network for predicting vibrational and phonon modes of molecules and periodic crystals, respectively. These predictions are made by evaluating the second derivative Hessian matrices of the learned energy model that is trained with the energy and force data.
Using this method, we are able to efficiently predict phonon dispersion and the density of states for inorganic crystal materials.
For molecules, we also derive the symmetry constraints for IR/Raman active modes by analyzing the phonon mode irreducible representations.
Additionally, we demonstrate that using Hessian as a new type of higher-order training data improves energy models beyond models that only use lower-order energy and force data.
With this second derivative approach, one can directly relate the energy models to the experimental observations for the vibrational properties.
This approach further connects to a broader class of physical observables with a generalized energy model that includes external fields.

\section{Introduction}
The phononic excitations in an atomic structure, i.e. the quantizations of the vibration modes~\cite{kittel2015introduction,Marder2010}, play a crucial role in determining the material properties such as mechanical, thermal, ferroelectricity, and transport characteristics in various applications~\cite{lowk_MgAgSb,phonon_glass,Phonovoltaic,Phonon_catalysis,Topological_phonon, vib_spectra_thermal, phonon_ferroelectric}.
In the Bardeen–Cooper–Schrieffer type of superconductivity~\cite{SC_BCS}, the phonons are essential in providing effective attractive interactions between electrons for forming Cooper pairs~\cite{SC_isotope1,SC_isotope2}.
In solids, unstable phonon modes at negative (or imaginary) energy indicate energetically favorable lattice distortions that lead to symmetry-lowering structural phase transitions, which are commonly found in, for example, perovskite structures~\cite{perovskite_phonon} and phase transitions~\cite{phonon_phase_transition}.
Due to the perturbative nature of the phonon modes, these vibrational modes often lower the symmetries of the structure~\cite{MDresselhaus_group_book}, which can also be viewed from the oscillating Bloch phases in describing the atomic displacements~\cite{kittel2015introduction}.
The symmetry characteristics of these modes further determine their behavior when they are studied using infrared (IR)/Raman spectroscopy~\cite{MDresselhaus_group_book}.
While conventional computational methods like density functional theory (DFT)~\cite{DFT_HK,DFT_Nobel,DFT_KS} and density functional perturbation theory(DFPT)~\cite{DFPT_Baroni,DFPT_phonon} calculations excel in predicting properties relevant to these applications, their computational expense makes them unsuitable for efficiently screening the extensive pool of potential material candidates.

Machine learning (ML) models~\cite{MLFF_Unke2021,ML_solids_Schmidt2019,ML_materials_Mobarak2023} offer a promising avenue for creating accurate and efficient predictive models for materials design. 
ML models trained on large-scale high-throughput databases of DFT calculation can make accelerated predictions of crucial information such the calculations for potential energy landscapes, atomic forces, and molecular dynamics (MD) simulations~\cite{MLPot_Behler,MLPot_gaussian,ML_MD,MLPot_physnet,MLPot_schnet}.
This is facilitated by the significant progress in parametrizing atomic potentials using physics-informed equivariant neural network models~\cite{e3nn_paper,Allegro,Batatia2022MACE,NequIP_paper,equiformer,equiformer_v2,escn}, especially relevant for MD simulations~\cite{jaxmd2020, MD_protein} and structural optimizations and relaxations.
These predictions expedite finding the equilibrium structure of trial material, thereby streamlining the process of screening and refining candidate crystal structures~\cite{Materials_Project}. 
Furthermore, ML models capacity can be applied to systems significantly larger in size compared to conventional methods (e.g. DFT) that are used to generate training data~\cite{Allegro}. 
This enables the use of prediction models to investigate the complex interaction for large atomic structures and molecules, relevant for drug discovery~\cite{Allegro,MLFF_drugs}.

In the high-throughput material search and AI-guided design framework, it is therefore crucial to have an efficient tool to predict the phonon properties given an atomic structure.
In this work, we utilize such symmetry-aware E(3)-equivariant graph neural network models~\cite{e3nn,e3nn_paper} to achieve efficient and accurate phonon predictions by direct computations of the Hessian matrices based on the machine learning potential models.
In contrast to the typical data types in modeling atomic structures, such as the atomic positions and types, phonons are more complicated and often visualized as phonon band structures.
These are denoted by the functions $\omega^n(\vec{k})$, which describe the energy of vibrational modes of a crystal with respect to crystal momentum $\vec{k}$ from the Bloch theorem~\cite{Bloch_th,kittel2015introduction,Marder2010}.
These functions have many symmetry properties: They are invariant under the shift with reciprocal lattice vectors $\vec{k} \rightarrow \vec{k} +\vec{G}$, and the basic unit can be captured by the function values within the Brillouin zone. 
Furthermore, crystallographic symmetries enforce strict constraints and relations on the $\omega^n(\vec{k})$ functions, tied to the ``little group'' symmetry representations in the momentum space~\cite{MDresselhaus_group_book}. 
The translational and rotational invariances of the lattice further lead to the acoustic sum rules~\cite{phonon_asr}.
These constraints give rise to specific band degeneracy patterns at high symmetry points and energy dispersions away from these points from their symmetry irreducible representations~\cite{MDresselhaus_group_book}.
For inorganic 3D periodic lattice crystals, there are 230 distinct types of the space groups for the crystallographic symmetries. 
It is therefore challenging to hard-code these symmetry features and constraints for the phonon band structure data. 
Previously, phonon predictions were shown to be derived from atomic potential models by evaluating the forces for the slightly displaced atomic patterns near their equilibrium positions~\cite{universalIAP,Owen2023TM23}. 
These methods require choices of supercell geometries and the enumerations of the independent displacement pattern subject to the crystallographic symmetry constraints, and the finite-difference approximation for getting numerical Hessians.
Due to the supercell geometries adopted, the phonon bands are folded and one has to apply band structure unfolding to get the conventional phonon spectrum~\cite{phonopy,phonopy-phono3py-JPSJ}.
On the other hand, direct dynamical matrix predictions such as the virtual node method VGNN~\cite{VGNNarxiv} give phonon predictions without generating an energy model from the training process (However, the additional symmetry constraints might have to be imposed manually to restore the acoustic sum rule).
In this work, we address how to represent the ``data type'' of a phonon band structure with a graph neural network in a way that captures the physical properties and symmetries naturally, how to evaluate the phonons directly from energy functions, and the duality between real-space and momentum space descriptions.
One advantage of our approach is the preservation of the relevant crystalline symmetries by using the equivariant graph neural network architecture, which automatically retains the symmetries in the phonon bands without additional work.

Besides the prediction for vibrational properties of an atomic structure, our training paradigm further generalizes the training data types by including the Hessian data as a higher-order extension beyond the energy and force data at the zeroth and first order respectively.
This ability is especially useful as it opens the possibility to use experimental data from measuring the vibrational properties to fine-tune the energy model. 
This opens a path to further improve the predictions of the energy model by correcting approximation errors in the simulated energy and force training data with the direct use of experimental data, rather than an indirect approach using experimental observable and statistics~\cite{trajectory_reweighting}.


\section{Results}
\subsection{Phonon formulations}

The physics of phonons can be viewed naturally from the local expansion of the potential energy landscape of an atomic structure~\cite{kittel2015introduction,Marder2010}.
While the zeroth order term gives the structural energy, the first order derivative terms, which determine the atomic forces, vanish for a locally equilibrium structure.
This leads to the next leading term in expanding the potential energy functional, the second derivative Hessians.

\begin{align}
\label{energy_expansion}
    E((\vec{x}_{ai})_{\alpha}) & \approx  E_0 + \\ \nonumber
    & \sum_{(\vec{x}_{ai})_{\alpha},(\vec{x}_{bj})_{\beta}} \frac{1}{2} \frac{\partial^2 E}{\partial (\vec{x}_{ai})_{\alpha} \partial (\vec{x}_{bj})_{\beta}}  (\vec{x}_{ai})_{\alpha}  (\vec{x}_{bj})_{\beta}
\end{align} for atomic coordinates $(\vec{x}_{ai})_{\alpha}$ in the structure.

These second-order terms are called force constants and are related to the stability of a crystal structure. For an equilibrium structure, the constants distinguish whether a structure is locally stable or at a saddle point in the energy landscape. 

Most relevant to this work, the force constants can be used to derive the phonon equation of motion of a periodic structure via constructing the dynamical matrix, defined as 
\begin{equation}
\label{phonon_dynmat}
    D_{\alpha \beta}(ij,\vec{q}) = \frac{1}{\sqrt{m_i m_j}} \sum_{a} \frac{\partial^2 E}{\partial (\vec{x}_{0i})_{\alpha} \partial (\vec{x}_{aj})_{\beta}} e^{i \vec{q} \cdot (\vec{x}_{aj} - \vec{x}_{0i})},
\end{equation} with $E$ the energy functional which can be described by a machine learning interatomic potential, the atomic coordinates $\vec{x}_{aj}$ with the unit cell index $a$ and cartesian coordinate $j$, and the atomic masses $m_i$. 

Given the dynamical matrix, the phonon eigenvalue equation is derived as
\begin{equation}
\label{phonon_eig}
    \sum_{j\beta} D_{\alpha \beta}(ij,\vec{q}) e^n_{\beta}(j,\vec{q}) = [\omega^n(\vec{q})]^2 e^n_{\alpha} (i,\vec{q})
\end{equation} for the quantized vibrational energies $\omega^{n}(\vec{q})$ and modes $e^n_{\alpha} (i,\vec{q})$.
Here we like to highlight the duality picture in Eq. \ref{phonon_dynmat}.
Experimentally, these vibrational modes can be probed by optical infrared and Raman spectroscopic techniques. Due to the optical nature ($\vec{q} \rightarrow 0$), these probes are usually limited to zero momentum states, with additional symmetry requirements to be optically active. Phonon bands can instead be probed by inelastic neutron and X-ray scattering measurements that map out the momentum-dependent phonon mode energies.

The above formulation informs our approach for predicting vibrational properties using equivariant energy models. Rather than predict a continuous function over momentum space directly, we can instead predict real space force constants which can be computed as second derivatives with respect to positions in our extended graph, without any loss of generality. Additionally, phonon modes are heavily constrained by symmetry, e.g. specific modes are required to cross at specific high symmetry $\vec{k}$. As long as an equivariant energy model is used, our derivative based formulation guarantees that our predictions will have the appropriate symmetry even before training. This means the model effectively only needs to learn relative energies; the only missing piece that symmetry cannot directly tell us (Ch. 8 and Ch. 11 in~\cite{MDresselhaus_group_book}). 

This equips us with the ability to rigorously associate our phonon modes with the irreducible representations which used to rigorously determine the ``selection rules'' for which modes will and will not be accessible via specific spectroscopies (e.g. modes that transform as vectors can be probed with infrared spectroscopies, but modes that transform as symmetric matrices cannot) (Ch. 8 in ~\cite{MDresselhaus_group_book}).
The ability to use real space quantities to derive a function over momentum space can be generalized to other physical quantities (such as the electronic bands and localized Wannier functions~\cite{wannier_review}). 


If a structure is at a local energy minimum, the phonons describe the restoring forces atoms experience when from their equilibrium positions. 
However if a structure is instead at a saddle point, the phonons can also indicate collective displacements that lower the energy of the structure.
These ``unstable'' phonon modes can appear in the phonon spectrum as modes with negative (or imaginary) mode frequencies. 
Unstable modes occur in crystal structures that experience structural phase transitions ~\cite{phonon_phase_transition}) such as perovskites crystals many of which distort to lower symmetry structures with a lowering of temperature ~\cite{perovskite_phonon}. 
These structural distortions can then be coupled to the electronic properties and provide another tuning knob such as the band reconstruction from the charge density waves~\cite{CsV3Sb5_CO}.

At long range $(\vec{q} \rightarrow \Gamma)$, the polarization from the atomic deformation pattern can generate an electric field that in turn modifies the phonon dispersion at long wavelength~\cite{Gonze_Phonon_theory,Phonon_NAC}. 
This correction depends on the $\vec{q}$ direction, and can lead to discontinuity in the phonon spectrum near $\Gamma$ point, also known as the LO-TO splitting~\cite{Gonze_Phonon_theory}. 
In our work, we focus on the short-range mechanical energy part in the dynamical matrix response. 
However, these dipole-dipole correction terms can be added to the dynamical matrix for the phonon equation of motion, when supplied with the additional information on the Bohr effective charges and the dielectric permittivity tensor from the DFPT calculations~\cite{Gonze_Phonon_theory}. 
To evaluate such corrections, one needs to construct an equivariant neural network to predict the Bohr effective charges $Z^*$~\cite{ion_migration} and the tensorial dielectric permittivity for atomic structures.
These physical quantities are relevant for many material applications, such as the ferroelectrics~\cite{Born_ferroelectric} and ion migration~\cite{ion_migration}, and we leave the modeling of these quantities to the future work (see supplementary for prototype equivariant models and the code).

\subsection{Phonax: a JAX framework for phonons}
3D Euclidean symmetry-equivariant graph neural network atomic potential models are built to preserve the symmetries of 3D data: 3D translation, rotation and inversion. This makes them data-efficient in predicting accurate atomic energy and force interactions of 3D atomistic systems. Examples of equivariant graph neural network atomic potential models include, NequIP~\cite{NequIP_paper}, MACE~\cite{Batatia2022MACE}, Allegro~\cite{Allegro}, etc.

The inputs to these models are 3D coordinates and atom types and the outputs are energies, which are invariant under 3D Euclidean symmetry. Between the inputs and outputs of these models, hidden representations are expressed as equivariant features that transform predictably when ``rotated''. In \texttt{e3nn}~\cite{e3nn,e3nn_paper} these equivariant features are broken into fundamental irreducible representation that are indexed by their angular frequency $L$, which transform according to their symmetry transformation matrices.


In our work, we have used NequIP~\cite{NequIP_paper} (implementation within JAX-MD~\cite{jaxmd2020}) and MACE~\cite{Batatia2022Design,Batatia2022MACE} models as our equivariant models for the interatomic energy functional $E$.
To evaluate the phonon dynamical matrix, we evaluate the second derivatives with respect to the pairs of atomic coordinates in the atomic structure as appear in the dynamical matrix in Eq. \ref{phonon_dynmat}, which is carried out by the automatic differentiation implemented in the JAX numerical framework~\cite{jax2018github}. 
As phonon modes can break the translation symmetries of the crystal structure, 
we have to generalize the conventional periodic crystal graph traditionally used for machine learning on crystal graphs. The original periodic graph only contains coordinates within a single unit cell, equivalent to $\vec{q} =0$ restriction of phonon modes.
Instead, we construct an extended graph for a periodic crystal, determined by the receptive field of the center unit cell in the message-passing graph neural network architecture.
All atomic sites within the receptive field are included and treated as independent variables in computing derivatives.
The size of the extended graph would grow with increasing $r_c$ cutoff radius and the number of steps $n_m$ for the message-passing (See supplementary section for examples).
In addition, we have implemented these derivations in phonax using using e3nn-jax~\cite{e3nn_jax,e3nn_paper}, jraph~\cite{jraph2020github}, and Haiku~\cite{haiku2020github} libraries.
Our phonax framework does not assume specific architectures for the ML potentials;
we only require that the machine learning potential architecture be equivariant and consistent with the extended graph construction used for computation.

\subsection{Experiments}
We assess the quality of the computational second-derivative Hessian matrices derived from phonax for the phonon and vibrational properties using various numerical experiments. 
We include periodic crystal and molecular cases.
We consider many ways to utilize Hessian predictions.
In Sections \ref{sec:phonon_lat}, \ref{sec:phonon_gen_inorg}, 
 and \ref{sec:mol_hess_pred} we considered material examples to predict the Hessian matrix and derive the vibrational properties using the converged energy models. 
In Sections \ref{sec:phonon_lat}, \ref{sec:data_aug_hess}, and \ref{sec:mol_spec_train} we evaluate the efficacy of Hessians used as training data treated to augment or fine-tune energy model training.
While the energy and force data used to train an energy model cannot be directly measured, the vibrational spectra can be directly compared to experimental probes. In Sections \ref{sec:mol_spec_train} and \ref{sec:mol_ir}, 
we compare the predicted spectra from phonax with DFT simulated spectra data and demonstrate that phonax can enable direct fine-tuning energy models on experimental spectra. 

Finally in Section \ref{sec:gen_field}, we point to future opportunities phonax presents. While in this work, we focus on Hessians of atomic coordinate, our framework is extendable to other derivatives 
such as those involving an external electric field with a generalized potential model. This would enable additional direct connections to physical observables.

\subsubsection{Phonon predictions for crystals with a periodic lattice}
\label{sec:phonon_lat}
\begin{figure}[h]
  \centering
  \includegraphics[width=\textwidth]{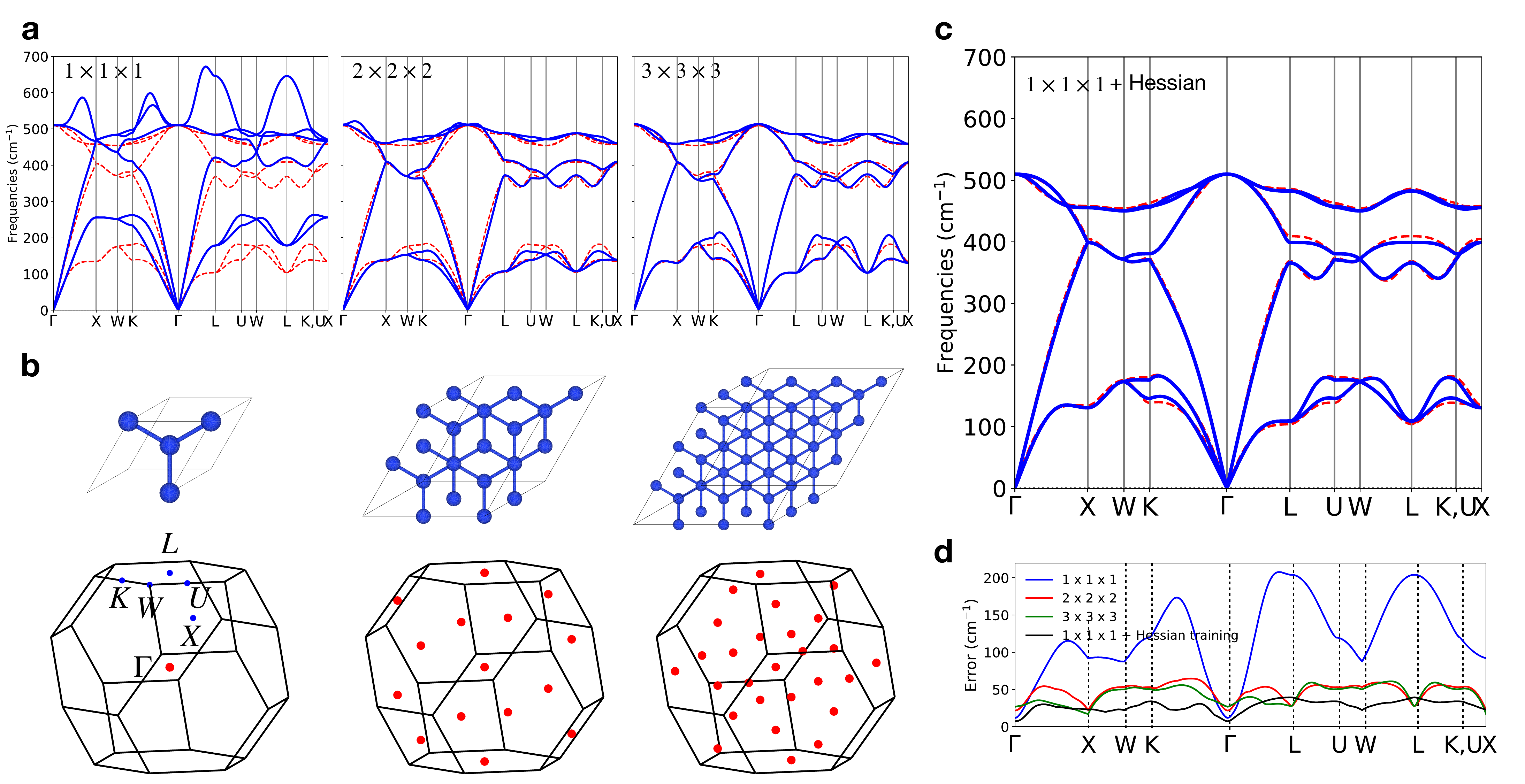}
  \caption{\label{Fig1}\textbf{Phonon prediction and training for silicon crystal} (a) Blue solid lines show the silicon crystal phonon predictions (mp-149) using the NequIP models trained with $1\!\times\!1\!\times\!1$, $2\!\times\!2\!\times\!2$ and $3\!\times\!3\!\times\!3$ supercell energy and force data with DFT phonon bands in red dashed lines. (b) These supercell geometries have 2, 16, 54 atoms respectively, and increasing sizes lead to finer mesh grid interpolations in the momentum space shown. (c) The energy model is trained with $1\!\times\!1\!\times\!1$ and the Hessian data. The phonon prediction is significantly better than using only $1\!\times\!1\!\times\!1$ data shown in a. (d) The error analysis in the momentum space for phonon predictions in a and c. At each $\vec{q}$, these errors are computed from the RMSE for the Hessian errors in the predicted dynamical matrix. The model trained with additional Hessian data (black) achieves the lowest error overall.}
\end{figure}

The most common DFT approach to obtain accurate phonon spectra uses  supercell geometries to calculate atomic interactions~\cite{phonopy}, and then derive the force constants from the finite-difference method or DFPT approach~\cite{DFPT_phonon}.
Intuitively, the supercell geometries have to be large enough to capture the decaying force constant dependence (see supplementary) to model the vibrational modes properly, which also captures the translational symmetry breaking from phononic vibrations.
Analogously, in Fig. \ref{Fig1}, we first show the predicted phonon spectrum for the silicon crystal in a diamond structure (mp-149 in the Materials Project~\cite{Materials_Project,phonon_DDB_data}) trained on increasingly larger supercell force data.
We trained the NequIP energy model~\cite{NequIP_paper} using energy and force data from various sizes of supercells in Fig. \ref{Fig1}a from smallest primitive $1\!\times\!1\!\times\!1$ unit cell to the largest $3\!\times\!3\!\times\!3$ supercell unit (see methods), and the phonon predictions are more accurate for larger supercells to allow independent degrees of freedom in perturbing the atomic coordinates.
From the reciprocal momentum space, larger supercells can be viewed as the finer interpolations of the Brillouin zone.
In Fig. \ref{Fig1}b, we show these unfolded $\vec{k}$ points for the supercell structures (in red), which are translationally invariant under the supercell lattice translations.
Alternatively, we augmented the smallest $1\!\times\!1\!\times\!1$ force data with Hessian data (see methods). Interestingly, these predictions outperformed those from the model trained on the largest $3\!\times\!3\!\times\!3$ supercell, which can be seen in Figs. \ref{Fig1}c and \ref{Fig1}d for the predicted spectrum and error analysis.
Instead of augmenting the training data with supercell energy and force data, the Hessian data can also improve the model by providing mutual pair interaction force constants that constrain the second-order derivatives of the energy landscape.
Experimentally, such phonon bands can be probed by inelastic neutron scattering measurements~\cite{Si_INS}. Having predicted the phonon mode energies here, our phonax framework enables the comparison with such experimental data, and backpropagate the correction signals to improve the underlying energy model, as we demonstrate in the molecular cases.

It is worth noting here that there are three $E=0$ modes at $\Gamma$ point. 
These zero modes appear due to the crystal energy invariance under a uniform translation of atomic positions and the resulting acoustic sum rule~\cite{phonon_asr}.
In our phonax framework, those zero modes naturally emerge from the inductive bias of the equivariant graph neural network model architecture~\cite{e3nn_paper}.
Had our model not have naturally handled this, these correction terms correction terms would be needed to restore these zero modes~\cite{Gonze_Phonon_theory}, as per the acoustic phonon sum rule.
In this example, we have demonstrated our phonax phonon prediction implementation, which also enables us to fine-tune the energy model with such second-order Hessian data.

\subsubsection{Phonon predictions for general inorganic crystal structures}
\label{sec:phonon_gen_inorg}
Having verified our phonon prediction method above for a single compound, we can further expand the phonon prediction capability by training energy models for crystals of generic types and structures, which would greatly facilitate the high-throughput searches for materials with specific phononic properties.
To converge the energy models, the training dataset has to include out-of-equilibrium structures with non-zero forces as well beyond the equilibrium structutres with vanishing forces.
To pursue such modeling, we considered the universal interatomic potentials dataset (universal-IAP, derived from the non-equilibrium structures in generating the Materials Project database and contains various forms and structures of compounds)~\cite{universalIAP} to train the energy model based on NequIP architecture~\cite{NequIP_paper} and MACE~\cite{Batatia2022MACE,Batatia2022Design} architectures.
With these models, we were able to make phonon predictions for more general types of crystal structures (See supplementary for the model hyper-parameters and phonon prediction examples).


However, to have a more rigorous phonon predictions comparison with the phonon band structures provided in the Materials Project~\cite{Materials_Project,phonon_DDB_data}, we had to generate our own energy and force DFT training dataset for near equilibrium crystal structures.
This is due to the differences in the DFT calculations employed in the universal-IAP dataset~\cite{universalIAP} (derived with Perdew Burke Ernzerhof (PBE)~\cite{pbe} exchange-correlation functional), and the phonon database~\cite{phonon_DDB_data} (computed with BPEsol exchange-correlation functional~\cite{VASP-PBEsol}).
Different functional choices lead to different predictions of equilibrium structures, atomic forces, and the resulting second-derivative Hessians that affect the phonon predictions.
To have a consistent energy and force training dataset, we carried out additional DFT calculations~\cite{vasp1,vasp2} with PBEsol~\cite{VASP-PBEsol} exchange-correlational functional to generate the energy and force data of the perturbed primitive unit cell and supercell structures, derived from the phonon database crystals (see methods for the dataset generation and the train-test split).

\begin{figure}[h]
  \centering
  \includegraphics[width=1.0\textwidth]{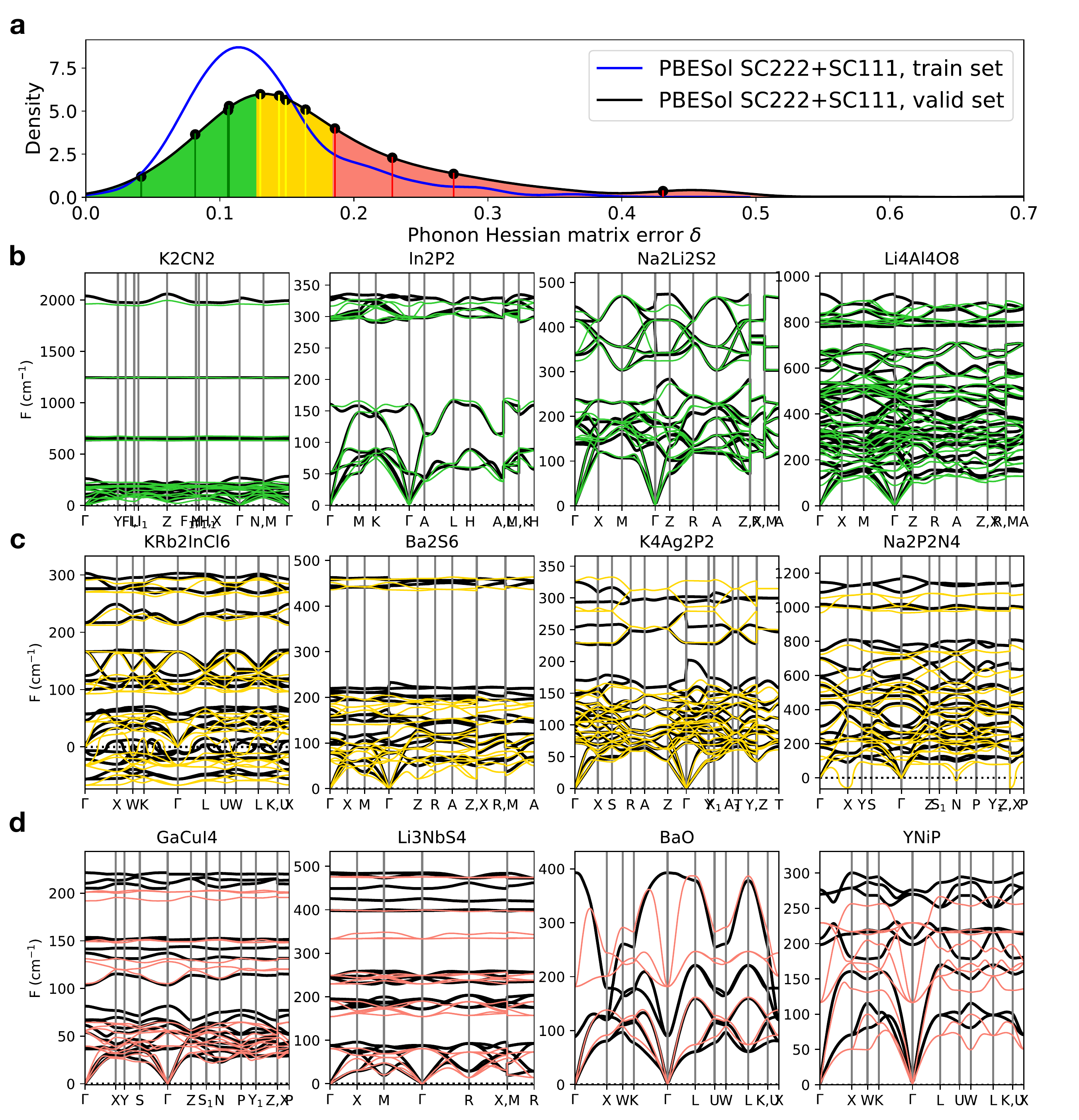}
  \caption{\label{Fig2}\textbf{Phonon prediction for general inorganic crystals with periodic lattices and the error analysis} (a) The distribution for the phonon predictions error metric of the periodic crystals evaluated for the training and valid datasets. The unseen valid crystals are grouped into three parts depending on the error metric (shaded areas). Crystal examples are drawn randomly from the (b) low (c) medium (d) high error groups in the prediction accuracy. The predicted phonon bands are colored green, yellow, and red respectively with the DFT results in black for comparison.}
\end{figure}

To gauge the phonon prediction accuracy, we adopted a relative error metric to measure the phonon prediction quality by comparing the predicted and target phononic dynamical matrices (see methods for details).
In these comparisons, we neglect the long-wavelength LO-TO splitting dipole-dipole corrections~\cite{Gonze_Phonon_theory} in the predicted dynamical matrix.
In Fig. \ref{Fig2}a, we performed this error analysis of the predicted phonon dynamical matrices based on the converged NequIP model~\cite{NequIP_paper} for the training and valid phonon datasets~\cite{phonon_DDB_data}, and classified the crystalline solids in the valid dataset into three parts depending on their errors (low, medium, and high error).
The model shown here is trained using only the PBEsol dataset we generated, without additional pre-training based on the universal-IAP dataset (see supplemetary for comparisons with other training procedures with and without such universal-IAP pre-training step.).
The shift in the error distributions reflects the errors in the phonon predictions for crystals used in the training dataset, versus the unseen ones in the valid dataset.
Random examples are drawn from each error population to showcase the phonon predictions, in Fig. \ref{Fig2} b-d.
Note even for the weaker phonon band predictions (red), the predicted phonon bands are still physically meaningful with symmetry features preserved.
In realistic applications for materials, we expect the predictions can be improved by adding more training data with specific crystal types or structures that are tailored for specific material applications.
As a material application example, we further predict the crystal structures that can host flat phonon bands in the spectrum, which are shown to be relevant for ferroelectric applications~\cite{flat_phonon_ferroelectric} (see supplementary).
With the correlation effects from the flat band electrons, it is intriguing to consider the electron-phonon couplings, and their implications on the superconductivity~\cite{SC_BCS} and structural instabilities.

\subsubsection{Molecular Hessian predictions}
\label{sec:mol_hess_pred}
The ability to predict molecular vibrational modes is crucial in experimental characterization.
Measurements such as infrared / Raman spectroscopy and inelastic neutron scattering (INS) probe the vibration modes of the molecular structures~\cite{MDresselhaus_group_book}.
Experimentally, these spectroscopies can be viewed as the fingerprints to identify the crystal structures and material quality.
Datasets constructed from the simulated INS spectra can further facilitate the high-throughput searches and discoveries for complex atomic structures~\cite{INS_dataset}.
It would be beneficial to predict the molecular vibrational modes and their symmetry properties based on accurate interatomic potential models as in the periodic crystalline solids discussed above, which would be relevant for chemical and pharmaceutical applications.
To achieve this, we have pre-trained an energy model for molecules (see methods).
Our phonax Hessian derivations can then be directly applied to these molecular cases as well, with one simplification: the evaluations for the Hessian matrix can be done without the extended graph construction due to the lack of periodic boundary conditions and repeated images.
By diagonalizing the dynamical matrix, we can derive the vibrational spectrum and normal modes. 
We can further analyze the symmetry irreducible representation of the modes at non-zero energies and identify the infrared/Raman active modes based on the symmetry selection rules (Ch. 8 in \cite{MDresselhaus_group_book}), with posym library~\cite{posym} and point group algebra (See methods and supplementary for the CH$_4$ molecular example).
These vibrational modes and optical spectroscopy predictions can yield insights for experimental molecular characterizations.
Beyond the binary predictions for the phononic Raman activity, one can further consider extensions to predict the full Raman tensors~\cite{MDresselhaus_group_book,Raman_tensor} that gives more details to the Raman responses, which is left to the future work.

\subsubsection{Training and data augmentation with molecular Hessians}
\label{sec:data_aug_hess}
\begin{figure}[h]
  \centering
  \includegraphics[width=0.8\textwidth]{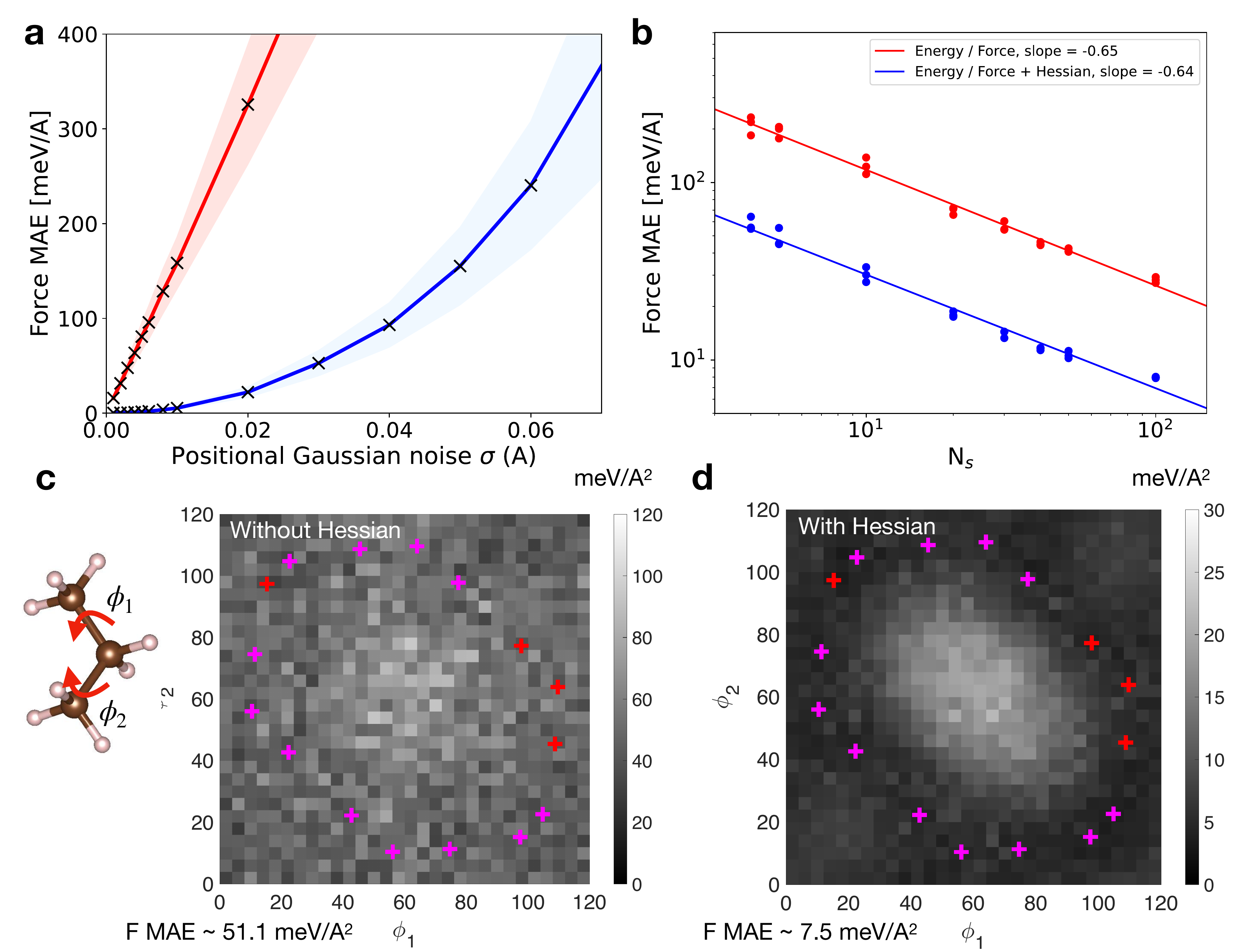}
  \caption{\label{Fig3}\textbf{Energy model improvements with augmented Hessian training data for molecules and the atomic force evaluations} (a) The evaluation for energy model trained with a single ethanol configuration, using only the energy/force data (red), versus training with additional Hessian data (blue). The force MAE is computed with slightly perturbed molecular configurations with positional Gaussian noise $\sigma$ from 0.001\AA\ to 0.07\AA\ from the trained configuration. 100 trainings (seeds + random samples) were used at each $\sigma$ with the force MAE std represented by the shaded area. With Hessian data training, force MAE is strongly suppressed around the molecular configuration. (b) The learning curve for the prediction error with the number of training configurations. The red (blue) curve shows the scaling from training energy and force data without (with) Hessians. The shift in curves indicates improvements from Hessian training. (c,d) To further demonstrate the training with Hessians, we consider the propane example described by two varying torsion angles. Four geometric configurations (sampled) are used for (c) energy force-only training and (d) energy, force and hessian training. The color plots show the force MAE map for slightly perturbed structure around the torsion angle. With Hessian training, the force errors are strongly suppressed. The red symbols indicate the randomly sampled configurations for the training dataset used with symmetry equivalent configurations in magenta.}
\end{figure}

So far, we have focused on the Hessian predictions from the converged energy models.
However, phonax framework also allows us to use Hessians differently as additional training data to improve the energy model.
We investigate how training with molecular Hessians affect the learning curve (see methods for the ethanol molecular dataset generated).
First in Fig. \ref{Fig3}a, we show how the force generalization errors are affected by a single molecular configuration training, with only its energy and force data, compared with the training utilizing energy, force and Hessian data simultaneously.
The force validation datasets are constructed by adding positional Gaussian noise at various strength to the training ethanol configuration sampled.
The scaling curve shows how the generalization errors are much reduced for locally perturbed structures with the additional Hessian training data (Note the scaling of errors changes from the linear scaling to the quadratic scaling with noise strength, which can be viewed intuitively from the Taylor's expansion).

Next in Fig. \ref{Fig3}b, we show the learning curve comparisons with and without using the hessian data for the force MAE in the valid dataset, with varying number of molecular training configurations used.
Although the force MAE improves when the Hessian data is included, the slope remains about the same, which is different from the scaling improvement between invariant and equivariant models~\cite{NequIP_paper}.
The shifted learning curve further suggests an effective $\sim 7$ times the number of force configurations in this training case with ethanol.
Physically, the Hessian provides a more complete prescription of the energy landscape around a given atomic structure compared to the energy and force-only training, and therefore effectively augments the training configurations and constrains the energy landscape.

We also investigate whether training on Hessian data improves predictions for the propane molecule.
A propane molecule, see Fig. \ref{Fig3}cd), can be described by its two torsion angles $\phi_1, \phi_2$ (periodic with $\phi_i \rightarrow \phi_i \pm 2\pi/3$ from the symmetry).
At each torsion angle configuration $(\phi_1,\phi_2)$, we generate a few randomly perturbed structures by adding Gaussian noise to the atomic positions (see methods for the dataset and training procedure).
The plots compare the force MAE when trained with only the energy and force data at randomly selected configurations in Fig. \ref{Fig3}c, and the training with additional Hessian data in Fig. \ref{Fig3}d (four selected training configurations and their symmetry equivalents are indicated in these maps).
The force MAE shows substantial improvements given the same number of configurations but additionally trained on Hessians, with the prediction errors much suppressed for structures near the sampled training configurations. Intuitively, this benefit stems from the Hessian constraining the local second derivative curvatures of the energy model.

Beyond the molecular Hessians, phonax can be applied to the periodic crystal cases as well to utilize crystal Hessians (as in the Si example) to improve the models.
It remains an interesting open question to see if the energy model trained with such additional Hessian data would be more stable under MD simulations~\cite{Force_not_enough}, as local stability is controlled by the second derivative Hessians.

\subsubsection{Molecular vibrational spectra training and energy model fine-tuning}
\label{sec:mol_spec_train}
The training of the energy models relies on the energy and forces datasets generated by the theoretical simulation methods such as the DFT~\cite{DFT_Nobel,DFT_HK,DFT_KS} calculations or the more advanced Ab initio quantum chemistry methods~\cite{CCSD1,CCSD2,Config_interact}.
Compared to the experimental data, these methods can be biased due to the unavoidable approximation errors and assumptions in the quantum many-body system.
Enabling the energy model to be trained or fine-tuned using the experimental measurements could eliminate such biases and errors in making predictions.
However, microscopic atomic energies and forces are not accessible to experimental measurements, and thus cannot be used to improve the training data quality.
Some have proposed indirect methods by using trajectory resampling~\cite{trajectory_reweighting} to compute the thermodynamic observables, which can then be compared with the experimental data and backpropogated to train the model.
It is therefore desirable to have a numerical approach that can utilize experimental data more directly to improve the energy model.

The second-derivative Hessians as derived in phonax can fill in the gap here as they are related to the vibrational properties of the structures which can be investigated experimentally via infrared, Raman, and neutron scattering probes as predicted above~\cite{MDresselhaus_group_book}.
In the previous section, full Hessian data as matrices is shown to improve the model training.
Here we consider using the eigenvalues instead, or the spectrum of the dynamical matrix, as the training target.
For our demonstration, we consider the DFT simulated molecular vibrational spectrum data as our target~\cite{INS_dataset}.
For specific molecular examples, we consider the benzene molecule as the training data for the fine-tuning process, toluene and phenol molecules as the test data, with a benzene ring structure and an additional functional group attached.
First, we use a pre-trained energy model to predict the vibrational spectrum for all three molecules (see methods).
Differences in the predicted and target vibrational spectrum are expected due to the different DFT settings employed in training the energy model, and the spectrum database.
For benzene, the difference in the spectrum, defined by the mean square errors in the sorted eigenvalue spectrum, can be used to define the loss function $\mathcal{L}$ in fine-tuning the model.
Once the energy model is fine-tuned for the benzene molecule based on the spectra comparison, we tested the fine-tuned model for the spectrum predictions of toluene and phenol molecules.
Without being trained to fine-tune on the toluene and phenol spectrum, the fine-tuned energy model shows improvement in the predictions in Fig. \ref{Fig4}b.
We argue that the corrections made to the energy model made during this training procedure can improve predictions for unseen molecules.

Beyond the demonstration with simulated spectrum data, we propose that our approach can be applied to the experimental spectrum data as well to fine-tune the energy for realistic data.
It remains an open question for the training procedure to utilize the various types of experimental data available such as optical spectroscopy and inelastic neutron scattering data, which can involve additional symmetry selection rules~\cite{MDresselhaus_group_book}, the matrix element effects, the training schedule, and how to define a proper loss function in comparing the frequency spectrums~\cite{spectra_opt_transport} that facilitate the training.
Although existing machine learning models can be used to predict the inelastic neutron scattering spectra~\cite{INS_ML}, these models do not yield a microscopic energy model that can be used for other downstream tasks.
The advantage of phonax is that it naturally connects the energy model with such experimental data types. 
In the next section, we further generalize such connections.

\subsubsection{Generalized energy model with external electric field for molecules}
\label{sec:gen_field}
\begin{figure}[h]
  \centering
  \includegraphics[width=1.0\textwidth]{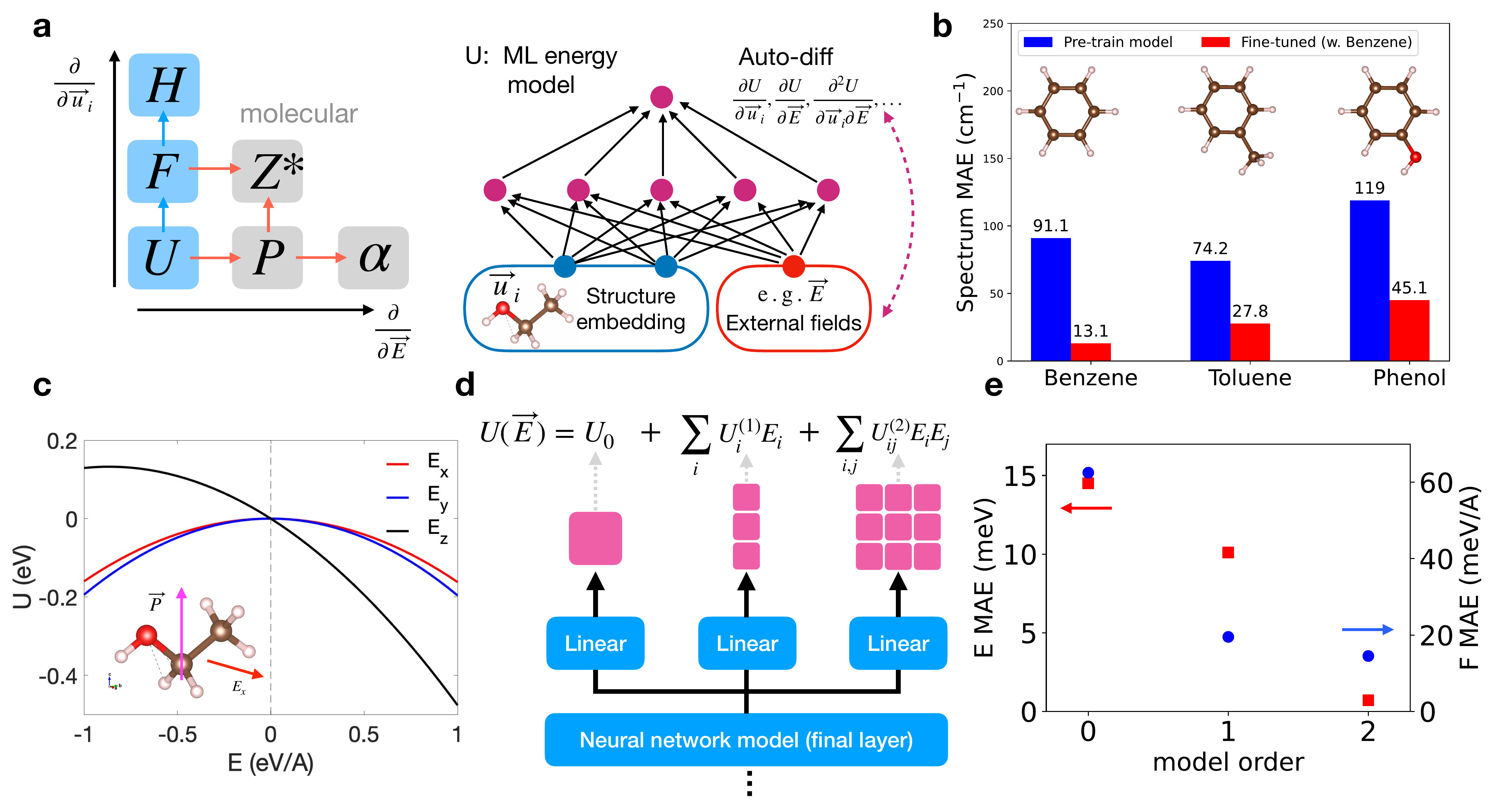}
  \caption{\label{Fig4}\textbf{Energy model generalizations with an external electric field, and the physical quantities from higher-order derivatives} (a) By generalizing the energy functional $U$ to include the electric field $\vec{E}$ dependence (besides the molecular graph variables such as the atomic coordinate $\vec{u}_i$ and atomic types $z_i$), one can derive the atomic forces $F$ and Hessians $H$, Born effective charge $Z^*$, polarization $P$, polarizability tensor $\alpha$ (molecular cases for simplicity) by taking the $\vec{u}_i$ and $\vec{E}$ derivatives as shown in the diagram via the auto-diff framework. These quantities can be related to various properties or physical observables such as phonon bands, spectroscopy, and molecular polarizations. (b) As an example, we consider the training with molecular vibrational spectrum. Given a pre-trained energy model, the energy model fine-tuned based on the benzene spectrum comparisons leads to improvements for toluene and phenol vibrational spectrum predictions. To demonstrate the energy model generalization with an electric field, (c) the total energy for an ethanol molecule with an electric field applied along (perpendicular) to the molecular electric polarization is shown in black (red and blue for the two perpendicular directions). These dependences support a perturbative expansion. (d) Generalized NequIP architecture~\cite{NequIP_paper} used to capture an electric field $\vec{E}$ dependence in the total energy. (e) The energy and force MAE for models expanded to higher $\vec{E}$ orders (the full model with order $= 2$ is shown in d).}
\end{figure}

The Hessian data we considered so far is the second derivative for the total energy functional with respect to the atomic positions $x_i$.
In this section, we further generalize our approaches beyond the second-order Hessian derivations to other broader scenarios that enable connections to various types of controlled parameters.
For example, if an external electric field is introduced to the energy model, one can obtain the electric polarizations $\vec{P}$ and polarizabilities $\alpha$ by taking derivatives with respect to the electric field as illustrated in \ref{Fig4}a.
When combined with mixed derivatives ($x_i$ and $\vec{E}$), one can further obtain quantities such as the Born effective charges $Z^*$.
These generalizations naturally relate the energy model to various physical observables and measurements.

For simplification, we focus on the molecular cases and consider an external electric field $\vec{E}$.
Due to the charge distribution and their movements, the electric field can induce changes in atomic energy and forces.
With this electric field $\vec{E}$ as an additional control parameter, physical quantities (molecular $\vec{P}$, $Z^*$ and $\alpha$ can be derived by taking the proper derivatives, as shown in \ref{Fig4}a diagram.
To demonstrate this, we generated an ethanol molecular dataset.
In Fig. \ref{Fig4}c, we show the energy variations when the electric field is applied along the electric polarization direction, or perpendicular to it.
We considered a generalized energy model that includes an electric field coupling up to the quadratic terms perturbatively (model order = 2 in this case, with a modified NequIP energy model architecture in \ref{Fig4}d, see methods).
In Fig. \ref{Fig4}e, we show that these generalized electric field terms improve the energy and force predictions for the molecules when increasing the model order.

Similarily, the Born effective charge $Z^{*}$ can also be predicted by evaluating the $\partial^2 U/\partial \vec{E} \partial \vec{x}_i$ via auto-differentiation as implemented in JAX~\cite{jax2018github}.
Besides the potentials of these generalized energy models in the electro-chemistry simulations and applications, this framework enables one to consider the coupling to externally controlled parameters more broadly (for example, magnetic field $H$, magnetization $M$, magnetic susceptibility $\chi$, and magnetoelectric effect~\cite{magnetoelectric}), their systematic expansions by perturbation orders, and the derivations for the relevant physical quantities.
From a machine learning perspective, this brings data of different physical sources and modalities into the same framework based on the general energy model (see also discussions in Ref. ~\cite{Universal_MLpot}).
As these quantities are different facets of the underlying quantum mechanical physics principles, it remains to be seen if training an energy model on these different modalities connected through this formalism leads to a stronger and more generalizable physical surrogate model, or 
as a better pre-trained model for downstream tasks with emergent abilities, as in large language models~\cite{LLM_emergent}.

\section{Discussion}
In this work, we introduced the phonax framework, a JAX framework that derives the second-derivative Hessians for phonons based on E(3)-equivariant graph neural network interatomic potentials.
There are two aspects of our work, and the first one is the efficient and accurate prediction of the vibrational characteristics of both periodic crystals and molecules. 
In a high-throughput search for the materials, this is a crucial step to screen or evaluate the material candidates in the applications, and downstream tasks involving thermal and transport properties~\cite{Phoebe}.
The general potential pre-trained models enable such phononic and stability predictions for crystals of various elemental and structural types.
Additionally, we showed that incorporating Hessian data or the spectrum eigenvalues into the input dataset enhances the training of energy models.
Intuitively, the vibrational phonon properties inform the energetic and mechanical properties of the materials.
Despite being the second-derivative data, their properties connect more directly to realistic experimental observations compared to the first-order (unobservable microscopic) force data and atomic energies.
The phonax prediction model would further allow the training or fine-tuning to use both experimental or simulated phonon data to improve the underlying energy model.
The connections between experimental observables and higher-order derivatives also seem natural with a generalized potential model, in which polarizations and polarizabilities can be derived when including the electric field dependence.
To further extend the model, one can consider the phonon couplings to other (low energy) degrees of freedom in the material, such as the electrons and magnon spin excitations~\cite{magnon_phonon_coupling}.
The possibility to control these excitations by mechanical phonons would have technological implications, which include the information processing and storage, and exotic quantum phases driven by these couplings.
From the machine learning perspective, our phonax model can be combined with other predictive models to investigate the phonon coupling effects by providing the vibrational properties.


\section{Code and data availability}
The phonax code and datasets will be made available at atomicarchitects group GitHub: https://github.com/atomicarchitects/phonax.
The generalized energy model and the predictive models for Born effective charge and dielectric constant will be made available at https://github.com/shiangfang/e3nn-models.


\section{Acknowledgement and Funding}
We appreciate fruitful discussions with Ilyes Batatia, Yi-Lun Liao, Aria Mansouri Tehrani, Max Shirokawa, Caolan John, Peter Miedaner, Nina Andrejevic, Maria Chan, Mingda Li, and Alexandre Tkatchenko.
This work was funded by the DOE ICDI grant DE-SC0022215 and the Gordon and Betty Moore Foundation EPiQS Initiative, Grant No. GBMF9070.
The authors acknowledge the MIT SuperCloud \cite{reuther2018interactive} and Lincoln Laboratory Supercomputing Center for providing HPC resources that have contributed to the results reported herein.


\section{Competing Interests Statement}
The authors declare no competing interests. 




\bibliography{references}

\begin{thebibliography}{10}
\expandafter\ifx\csname url\endcsname\relax
  \def\url#1{\texttt{#1}}\fi
\expandafter\ifx\csname urlprefix\endcsname\relax\def\urlprefix{URL }\fi
\providecommand{\bibinfo}[2]{#2}
\providecommand{\eprint}[2][]{\url{#2}}

\bibitem{kittel2015introduction}
\bibinfo{author}{Kittel, C.}, \bibinfo{author}{McEuen, P.} \&
  \bibinfo{author}{Sons, J. W.~.}
\newblock \emph{\bibinfo{title}{Introduction to Solid State Physics}}
  (\bibinfo{publisher}{John Wiley \& Sons}, \bibinfo{year}{2015}).
\newblock \urlprefix\url{https://books.google.com/books?id=rAMujwEACAAJ}.

\bibitem{Marder2010}
\bibinfo{author}{Marder, M.~P.}
\newblock \emph{\bibinfo{title}{Condensed Matter Physics}}
  (\bibinfo{publisher}{Wiley}, \bibinfo{year}{2010}).
\newblock \urlprefix\url{https://doi.org/10.1002/9780470949955}.

\bibitem{lowk_MgAgSb}
\bibinfo{author}{Li, X.} \emph{et~al.}
\newblock \bibinfo{title}{Ultralow thermal conductivity from transverse
  acoustic phonon suppression in distorted crystalline alpha-mgagsb}.
\newblock \emph{\bibinfo{journal}{Nature Communications}}
  \textbf{\bibinfo{volume}{11}}, \bibinfo{pages}{942} (\bibinfo{year}{2020}).

\bibitem{phonon_glass}
\bibinfo{author}{Takabatake, T.}, \bibinfo{author}{Suekuni, K.},
  \bibinfo{author}{Nakayama, T.} \& \bibinfo{author}{Kaneshita, E.}
\newblock \bibinfo{title}{Phonon-glass electron-crystal thermoelectric
  clathrates: Experiments and theory}.
\newblock \emph{\bibinfo{journal}{Rev. Mod. Phys.}}
  \textbf{\bibinfo{volume}{86}}, \bibinfo{pages}{669--716}
  (\bibinfo{year}{2014}).
\newblock \urlprefix\url{https://link.aps.org/doi/10.1103/RevModPhys.86.669}.

\bibitem{Phonovoltaic}
\bibinfo{author}{Melnick, C.} \& \bibinfo{author}{Kaviany, M.}
\newblock \bibinfo{title}{Phonovoltaic. i. harvesting hot optical phonons in a
  nanoscale p-n junction}.
\newblock \emph{\bibinfo{journal}{Phys. Rev. B}} \textbf{\bibinfo{volume}{93}},
  \bibinfo{pages}{094302} (\bibinfo{year}{2016}).
\newblock \urlprefix\url{https://link.aps.org/doi/10.1103/PhysRevB.93.094302}.

\bibitem{Phonon_catalysis}
\bibinfo{author}{Gordiz, K.}, \bibinfo{author}{Muy, S.},
  \bibinfo{author}{Zeier, W.~G.}, \bibinfo{author}{Shao-Horn, Y.} \&
  \bibinfo{author}{Henry, A.}
\newblock \bibinfo{title}{Enhancement of ion diffusion by targeted phonon
  excitation}.
\newblock \emph{\bibinfo{journal}{Cell Reports Physical Science}}
  \textbf{\bibinfo{volume}{2}}, \bibinfo{pages}{100431} (\bibinfo{year}{2021}).

\bibitem{Topological_phonon}
\bibinfo{author}{Peng, B.}, \bibinfo{author}{Hu, Y.},
  \bibinfo{author}{Murakami, S.}, \bibinfo{author}{Zhang, T.} \&
  \bibinfo{author}{Monserrat, B.}
\newblock \bibinfo{title}{Topological phonons in oxide perovskites controlled
  by light}.
\newblock \emph{\bibinfo{journal}{Science Advances}}
  \textbf{\bibinfo{volume}{6}} (\bibinfo{year}{2020}).
\newblock \urlprefix\url{https://doi.org/10.1126/sciadv.abd1618}.

\bibitem{vib_spectra_thermal}
\bibinfo{author}{Skelton, J.~M.} \emph{et~al.}
\newblock \bibinfo{title}{Lattice dynamics of the tin sulphides sns2{,} sns and
  sn2s3: vibrational spectra and thermal transport}.
\newblock \emph{\bibinfo{journal}{Phys. Chem. Chem. Phys.}}
  \textbf{\bibinfo{volume}{19}}, \bibinfo{pages}{12452--12465}
  (\bibinfo{year}{2017}).

\bibitem{phonon_ferroelectric}
\bibinfo{author}{Kamba, S.}
\newblock \bibinfo{title}{{Soft-mode spectroscopy of ferroelectrics and
  multiferroics: A review}}.
\newblock \emph{\bibinfo{journal}{APL Materials}} \textbf{\bibinfo{volume}{9}},
  \bibinfo{pages}{020704} (\bibinfo{year}{2021}).
\newblock \urlprefix\url{https://doi.org/10.1063/5.0036066}.
\newblock
  \eprint{https://pubs.aip.org/aip/apm/article-pdf/doi/10.1063/5.0036066/13781671/020704\_1\_online.pdf}.

\bibitem{SC_BCS}
\bibinfo{author}{Bardeen, J.}, \bibinfo{author}{Cooper, L.~N.} \&
  \bibinfo{author}{Schrieffer, J.~R.}
\newblock \bibinfo{title}{Theory of superconductivity}.
\newblock \emph{\bibinfo{journal}{Phys. Rev.}} \textbf{\bibinfo{volume}{108}},
  \bibinfo{pages}{1175--1204} (\bibinfo{year}{1957}).
\newblock \urlprefix\url{https://link.aps.org/doi/10.1103/PhysRev.108.1175}.

\bibitem{SC_isotope1}
\bibinfo{author}{Reynolds, C.~A.}, \bibinfo{author}{Serin, B.},
  \bibinfo{author}{Wright, W.~H.} \& \bibinfo{author}{Nesbitt, L.~B.}
\newblock \bibinfo{title}{Superconductivity of isotopes of mercury}.
\newblock \emph{\bibinfo{journal}{Phys. Rev.}} \textbf{\bibinfo{volume}{78}},
  \bibinfo{pages}{487--487} (\bibinfo{year}{1950}).
\newblock \urlprefix\url{https://link.aps.org/doi/10.1103/PhysRev.78.487}.

\bibitem{SC_isotope2}
\bibinfo{author}{Maxwell, E.}
\newblock \bibinfo{title}{Isotope effect in the superconductivity of mercury}.
\newblock \emph{\bibinfo{journal}{Phys. Rev.}} \textbf{\bibinfo{volume}{78}},
  \bibinfo{pages}{477--477} (\bibinfo{year}{1950}).
\newblock \urlprefix\url{https://link.aps.org/doi/10.1103/PhysRev.78.477}.

\bibitem{perovskite_phonon}
\bibinfo{author}{Pilania, G.} \& \bibinfo{author}{Lookman, T.}
\newblock \bibinfo{title}{Electronic structure and biaxial strain in
  ${\mathrm{rbhgf}}_{3}$ perovskite and hybrid improper ferroelectricity in
  (na,rb)hg2f6 and (k,rb)hg2f6 superlattices}.
\newblock \emph{\bibinfo{journal}{Phys. Rev. B}} \textbf{\bibinfo{volume}{90}},
  \bibinfo{pages}{115121} (\bibinfo{year}{2014}).
\newblock \urlprefix\url{https://link.aps.org/doi/10.1103/PhysRevB.90.115121}.

\bibitem{phonon_phase_transition}
\bibinfo{author}{Adams, D.~J.} \& \bibinfo{author}{Passerone, D.}
\newblock \bibinfo{title}{Insight into structural phase transitions from the
  decoupled anharmonic mode approximation}.
\newblock \emph{\bibinfo{journal}{Journal of Physics: Condensed Matter}}
  \textbf{\bibinfo{volume}{28}}, \bibinfo{pages}{305401}
  (\bibinfo{year}{2016}).

\bibitem{MDresselhaus_group_book}
 \emph{\bibinfo{title}{Group Theory}} (\bibinfo{publisher}{Springer Berlin
  Heidelberg}, \bibinfo{year}{2008}).
\newblock \urlprefix\url{https://doi.org/10.1007%2F978-3-540-32899-5}.

\bibitem{DFT_HK}
\bibinfo{author}{Hohenberg, P.} \& \bibinfo{author}{Kohn, W.}
\newblock \bibinfo{title}{Inhomogeneous electron gas}.
\newblock \emph{\bibinfo{journal}{Phys. Rev.}} \textbf{\bibinfo{volume}{136}},
  \bibinfo{pages}{B864--B871} (\bibinfo{year}{1964}).
\newblock \urlprefix\url{https://link.aps.org/doi/10.1103/PhysRev.136.B864}.

\bibitem{DFT_Nobel}
\bibinfo{author}{Kohn, W.}
\newblock \bibinfo{title}{Nobel lecture: Electronic structure of matter—wave
  functions and density functionals}.
\newblock \emph{\bibinfo{journal}{Reviews of Modern Physics}}
  \textbf{\bibinfo{volume}{71}}, \bibinfo{pages}{1253–1266}
  (\bibinfo{year}{1999}).
\newblock \urlprefix\url{http://dx.doi.org/10.1103/RevModPhys.71.1253}.

\bibitem{DFT_KS}
\bibinfo{author}{Kohn, W.} \& \bibinfo{author}{Sham, L.~J.}
\newblock \bibinfo{title}{Self-consistent equations including exchange and
  correlation effects}.
\newblock \emph{\bibinfo{journal}{Phys. Rev.}} \textbf{\bibinfo{volume}{140}},
  \bibinfo{pages}{A1133--A1138} (\bibinfo{year}{1965}).
\newblock \urlprefix\url{https://link.aps.org/doi/10.1103/PhysRev.140.A1133}.

\bibitem{DFPT_Baroni}
\bibinfo{author}{Baroni, S.}, \bibinfo{author}{Giannozzi, P.} \&
  \bibinfo{author}{Testa, A.}
\newblock \bibinfo{title}{Green’s-function approach to linear response in
  solids}.
\newblock \emph{\bibinfo{journal}{Physical Review Letters}}
  \textbf{\bibinfo{volume}{58}}, \bibinfo{pages}{1861–1864}
  (\bibinfo{year}{1987}).
\newblock \urlprefix\url{http://dx.doi.org/10.1103/PhysRevLett.58.1861}.

\bibitem{DFPT_phonon}
\bibinfo{author}{Baroni, S.}, \bibinfo{author}{de~Gironcoli, S.},
  \bibinfo{author}{Dal~Corso, A.} \& \bibinfo{author}{Giannozzi, P.}
\newblock \bibinfo{title}{Phonons and related crystal properties from
  density-functional perturbation theory}.
\newblock \emph{\bibinfo{journal}{Reviews of Modern Physics}}
  \textbf{\bibinfo{volume}{73}}, \bibinfo{pages}{515–562}
  (\bibinfo{year}{2001}).
\newblock \urlprefix\url{http://dx.doi.org/10.1103/RevModPhys.73.515}.

\bibitem{MLFF_Unke2021}
\bibinfo{author}{Unke, O.~T.} \emph{et~al.}
\newblock \bibinfo{title}{Machine learning force fields}.
\newblock \emph{\bibinfo{journal}{Chemical Reviews}}
  \textbf{\bibinfo{volume}{121}}, \bibinfo{pages}{10142–10186}
  (\bibinfo{year}{2021}).
\newblock \urlprefix\url{http://dx.doi.org/10.1021/acs.chemrev.0c01111}.

\bibitem{ML_solids_Schmidt2019}
\bibinfo{author}{Schmidt, J.}, \bibinfo{author}{Marques, M. R.~G.},
  \bibinfo{author}{Botti, S.} \& \bibinfo{author}{Marques, M. A.~L.}
\newblock \bibinfo{title}{Recent advances and applications of machine learning
  in solid-state materials science}.
\newblock \emph{\bibinfo{journal}{npj Computational Materials}}
  \textbf{\bibinfo{volume}{5}} (\bibinfo{year}{2019}).
\newblock \urlprefix\url{http://dx.doi.org/10.1038/s41524-019-0221-0}.

\bibitem{ML_materials_Mobarak2023}
\bibinfo{author}{Mobarak, M.~H.} \emph{et~al.}
\newblock \bibinfo{title}{Scope of machine learning in materials research—a
  review}.
\newblock \emph{\bibinfo{journal}{Applied Surface Science Advances}}
  \textbf{\bibinfo{volume}{18}}, \bibinfo{pages}{100523}
  (\bibinfo{year}{2023}).
\newblock \urlprefix\url{http://dx.doi.org/10.1016/j.apsadv.2023.100523}.

\bibitem{MLPot_Behler}
\bibinfo{author}{Behler, J.} \& \bibinfo{author}{Parrinello, M.}
\newblock \bibinfo{title}{Generalized neural-network representation of
  high-dimensional potential-energy surfaces}.
\newblock \emph{\bibinfo{journal}{Phys. Rev. Lett.}}
  \textbf{\bibinfo{volume}{98}}, \bibinfo{pages}{146401}
  (\bibinfo{year}{2007}).
\newblock
  \urlprefix\url{https://link.aps.org/doi/10.1103/PhysRevLett.98.146401}.

\bibitem{MLPot_gaussian}
\bibinfo{author}{Bart\'ok, A.~P.}, \bibinfo{author}{Payne, M.~C.},
  \bibinfo{author}{Kondor, R.} \& \bibinfo{author}{Cs\'anyi, G.}
\newblock \bibinfo{title}{Gaussian approximation potentials: The accuracy of
  quantum mechanics, without the electrons}.
\newblock \emph{\bibinfo{journal}{Phys. Rev. Lett.}}
  \textbf{\bibinfo{volume}{104}}, \bibinfo{pages}{136403}
  (\bibinfo{year}{2010}).
\newblock
  \urlprefix\url{https://link.aps.org/doi/10.1103/PhysRevLett.104.136403}.

\bibitem{ML_MD}
\bibinfo{author}{Zhang, L.}, \bibinfo{author}{Han, J.}, \bibinfo{author}{Wang,
  H.}, \bibinfo{author}{Car, R.} \& \bibinfo{author}{E, W.}
\newblock \bibinfo{title}{Deep potential molecular dynamics: A scalable model
  with the accuracy of quantum mechanics}.
\newblock \emph{\bibinfo{journal}{Phys. Rev. Lett.}}
  \textbf{\bibinfo{volume}{120}}, \bibinfo{pages}{143001}
  (\bibinfo{year}{2018}).
\newblock
  \urlprefix\url{https://link.aps.org/doi/10.1103/PhysRevLett.120.143001}.

\bibitem{MLPot_physnet}
\bibinfo{author}{Unke, O.~T.} \& \bibinfo{author}{Meuwly, M.}
\newblock \bibinfo{title}{Physnet: A neural network for predicting energies,
  forces, dipole moments, and partial charges}.
\newblock \emph{\bibinfo{journal}{Journal of Chemical Theory and Computation}}
  \textbf{\bibinfo{volume}{15}}, \bibinfo{pages}{3678--3693}
  (\bibinfo{year}{2019}).

\bibitem{MLPot_schnet}
\bibinfo{author}{Sch{\"u}tt, K.~T.} \emph{et~al.}
\newblock \bibinfo{title}{{SchNet}: a continuous-filter convolutional neural
  network for modeling quantum interactions}.
\newblock In \emph{\bibinfo{booktitle}{Proceedings of the 31st International
  Conference on Neural Information Processing Systems}}, NIPS'17,
  \bibinfo{pages}{992--1002} (\bibinfo{publisher}{Curran Associates Inc.},
  \bibinfo{address}{Red Hook, NY, USA}, \bibinfo{year}{2017}).

\bibitem{e3nn_paper}
\bibinfo{author}{Geiger, M.} \& \bibinfo{author}{Smidt, T.}
\newblock \bibinfo{title}{e3nn: Euclidean neural networks}
  (\bibinfo{year}{2022}).
\newblock \urlprefix\url{https://arxiv.org/abs/2207.09453}.

\bibitem{Allegro}
\bibinfo{author}{Musaelian, A.} \emph{et~al.}
\newblock \bibinfo{title}{Learning local equivariant representations for
  large-scale atomistic dynamics}.
\newblock \emph{\bibinfo{journal}{Nature Communications}}
  \textbf{\bibinfo{volume}{14}}, \bibinfo{pages}{579} (\bibinfo{year}{2023}).

\bibitem{Batatia2022MACE}
\bibinfo{author}{Batatia, I.}, \bibinfo{author}{Kov{\'a}cs, D.~P.},
  \bibinfo{author}{Simm, G. N.~C.}, \bibinfo{author}{Ortner, C.} \&
  \bibinfo{author}{Cs{\'a}nyi, G.}
\newblock \bibinfo{title}{Mace: Higher order equivariant message passing neural
  networks for fast and accurate force fields} (\bibinfo{year}{2022}).
\newblock \eprint{2206.07697}.

\bibitem{NequIP_paper}
\bibinfo{author}{Batzner, S.} \emph{et~al.}
\newblock \bibinfo{title}{E(3)-equivariant graph neural networks for
  data-efficient and accurate interatomic potentials}.
\newblock \emph{\bibinfo{journal}{Nature Communications}}
  \textbf{\bibinfo{volume}{13}}, \bibinfo{pages}{2453} (\bibinfo{year}{2022}).

\bibitem{equiformer}
\bibinfo{author}{Liao, Y.-L.} \& \bibinfo{author}{Smidt, T.}
\newblock \bibinfo{title}{{Equiformer: Equivariant Graph Attention Transformer
  for 3D Atomistic Graphs}}.
\newblock In \emph{\bibinfo{booktitle}{International Conference on Learning
  Representations (ICLR)}} (\bibinfo{year}{2023}).
\newblock \urlprefix\url{https://openreview.net/forum?id=KwmPfARgOTD}.

\bibitem{equiformer_v2}
\bibinfo{author}{Liao, Y.-L.}, \bibinfo{author}{Wood, B.},
  \bibinfo{author}{Das*, A.} \& \bibinfo{author}{Smidt*, T.}
\newblock \bibinfo{title}{Equiformerv2: Improved equivariant transformer for
  scaling to higher-degree representations}.
\newblock \emph{\bibinfo{journal}{arxiv preprint arxiv:2306.12059}}
  (\bibinfo{year}{2023}).

\bibitem{escn}
\bibinfo{author}{Passaro, S.} \& \bibinfo{author}{Zitnick, C.~L.}
\newblock \bibinfo{title}{{Reducing SO(3) Convolutions to SO(2) for Efficient
  Equivariant GNNs}}.
\newblock In \emph{\bibinfo{booktitle}{International Conference on Machine
  Learning (ICML)}} (\bibinfo{year}{2023}).

\bibitem{jaxmd2020}
\bibinfo{author}{Schoenholz, S.~S.} \& \bibinfo{author}{Cubuk, E.~D.}
\newblock \bibinfo{title}{Jax m.d. a framework for differentiable physics}.
\newblock In \emph{\bibinfo{booktitle}{Advances in Neural Information
  Processing Systems}}, vol.~\bibinfo{volume}{33} (\bibinfo{publisher}{Curran
  Associates, Inc.}, \bibinfo{year}{2020}).
\newblock
  \urlprefix\url{https://papers.nips.cc/paper/2020/file/83d3d4b6c9579515e1679aca8cbc8033-Paper.pdf}.

\bibitem{MD_protein}
\bibinfo{author}{Lindorff-Larsen, K.}, \bibinfo{author}{Piana, S.},
  \bibinfo{author}{Dror, R.~O.} \& \bibinfo{author}{Shaw, D.~E.}
\newblock \bibinfo{title}{How fast-folding proteins fold}.
\newblock \emph{\bibinfo{journal}{Science}} \textbf{\bibinfo{volume}{334}},
  \bibinfo{pages}{517--520} (\bibinfo{year}{2011}).

\bibitem{Materials_Project}
\bibinfo{author}{Jain, A.} \emph{et~al.}
\newblock \bibinfo{title}{Commentary: The materials project: A materials genome
  approach to accelerating materials innovation}.
\newblock \emph{\bibinfo{journal}{{APL} Materials}}
  \textbf{\bibinfo{volume}{1}} (\bibinfo{year}{2013}).
\newblock \urlprefix\url{https://doi.org/10.1063/1.4812323}.

\bibitem{MLFF_drugs}
\bibinfo{author}{Chen, M.} \emph{et~al.}
\newblock \bibinfo{title}{The emergence of machine learning force fields in
  drug design}.
\newblock \emph{\bibinfo{journal}{Medicinal Research Reviews}}
  \textbf{\bibinfo{volume}{n/a}}.

\bibitem{e3nn}
\bibinfo{author}{Geiger, M.} \emph{et~al.}
\newblock \bibinfo{title}{Euclidean neural networks: e3nn}
  (\bibinfo{year}{2022}).
\newblock \urlprefix\url{https://doi.org/10.5281/zenodo.6459381}.

\bibitem{Bloch_th}
\bibinfo{author}{Bloch, F.}
\newblock \bibinfo{title}{{\"U}ber die quantenmechanik der elektronen in
  kristallgittern}.
\newblock \emph{\bibinfo{journal}{Zeitschrift f{\"u}r Physik}}
  \textbf{\bibinfo{volume}{52}}, \bibinfo{pages}{555--600}
  (\bibinfo{year}{1929}).

\bibitem{phonon_asr}
\bibinfo{author}{Leibfried, G.} \& \bibinfo{author}{Ludwig, W.}
\newblock \bibinfo{title}{Theory of anharmonic effects in crystals}.
\newblock vol.~\bibinfo{volume}{12} of \emph{\bibinfo{series}{Solid State
  Physics}}, \bibinfo{pages}{275--444} (\bibinfo{publisher}{Academic Press},
  \bibinfo{year}{1961}).

\bibitem{universalIAP}
\bibinfo{author}{Chen, C.} \& \bibinfo{author}{Ong, S.~P.}
\newblock \bibinfo{title}{A universal graph deep learning interatomic potential
  for the periodic table}.
\newblock \emph{\bibinfo{journal}{Nature Computational Science}}
  \textbf{\bibinfo{volume}{2}}, \bibinfo{pages}{718--728}
  (\bibinfo{year}{2022}).

\bibitem{Owen2023TM23}
\bibinfo{author}{Owen, C.~J.} \emph{et~al.}
\newblock \bibinfo{title}{Complexity of many-body interactions in transition
  metals via machine-learned force fields from the tm23 data set}
  (\bibinfo{year}{2023}).
\newblock \urlprefix\url{https://arxiv.org/abs/2302.12993}.

\bibitem{phonopy}
\bibinfo{author}{Togo, A.} \& \bibinfo{author}{Tanaka, I.}
\newblock \bibinfo{title}{First principles phonon calculations in materials
  science}.
\newblock \emph{\bibinfo{journal}{Scr. Mater.}} \textbf{\bibinfo{volume}{108}},
  \bibinfo{pages}{1--5} (\bibinfo{year}{2015}).

\bibitem{phonopy-phono3py-JPSJ}
\bibinfo{author}{Togo, A.}
\newblock \bibinfo{title}{First-principles phonon calculations with phonopy and
  phono3py}.
\newblock \emph{\bibinfo{journal}{J. Phys. Soc. Jpn.}}
  \textbf{\bibinfo{volume}{92}}, \bibinfo{pages}{012001}
  (\bibinfo{year}{2023}).

\bibitem{VGNNarxiv}
\bibinfo{author}{Okabe, R.} \emph{et~al.}
\newblock \bibinfo{title}{Virtual node graph neural network for full phonon
  prediction} (\bibinfo{year}{2023}).
\newblock \eprint{arXiv:2301.02197}.

\bibitem{trajectory_reweighting}
\bibinfo{author}{Thaler, S.} \& \bibinfo{author}{Zavadlav, J.}
\newblock \bibinfo{title}{Learning neural network potentials from experimental
  data via differentiable trajectory reweighting}.
\newblock \emph{\bibinfo{journal}{Nature Communications}}
  \textbf{\bibinfo{volume}{12}}, \bibinfo{pages}{6884} (\bibinfo{year}{2021}).

\bibitem{wannier_review}
\bibinfo{author}{Marzari, N.}, \bibinfo{author}{Mostofi, A.~A.},
  \bibinfo{author}{Yates, J.~R.}, \bibinfo{author}{Souza, I.} \&
  \bibinfo{author}{Vanderbilt, D.}
\newblock \bibinfo{title}{Maximally localized {Wannier} functions: Theory and
  applications}.
\newblock \emph{\bibinfo{journal}{Rev. Mod. Phys.}}
  \textbf{\bibinfo{volume}{84}}, \bibinfo{pages}{1419--1475}
  (\bibinfo{year}{2012}).

\bibitem{CsV3Sb5_CO}
\bibinfo{author}{Kang, M.} \emph{et~al.}
\newblock \bibinfo{title}{Charge order landscape and competition with
  superconductivity in kagome metals}.
\newblock \emph{\bibinfo{journal}{Nature Materials}}
  \textbf{\bibinfo{volume}{22}}, \bibinfo{pages}{186--193}
  (\bibinfo{year}{2023}).

\bibitem{Gonze_Phonon_theory}
\bibinfo{author}{Gonze, X.} \& \bibinfo{author}{Lee, C.}
\newblock \bibinfo{title}{Dynamical matrices, born effective charges,
  dielectric permittivity tensors, and interatomic force constants from
  density-functional perturbation theory}.
\newblock \emph{\bibinfo{journal}{Phys. Rev. B}} \textbf{\bibinfo{volume}{55}},
  \bibinfo{pages}{10355--10368} (\bibinfo{year}{1997}).
\newblock \urlprefix\url{https://link.aps.org/doi/10.1103/PhysRevB.55.10355}.

\bibitem{Phonon_NAC}
\bibinfo{author}{Pick, R.~M.}, \bibinfo{author}{Cohen, M.~H.} \&
  \bibinfo{author}{Martin, R.~M.}
\newblock \bibinfo{title}{Microscopic theory of force constants in the
  adiabatic approximation}.
\newblock \emph{\bibinfo{journal}{Phys. Rev. B}} \textbf{\bibinfo{volume}{1}},
  \bibinfo{pages}{910--920} (\bibinfo{year}{1970}).
\newblock \urlprefix\url{https://link.aps.org/doi/10.1103/PhysRevB.1.910}.

\bibitem{ion_migration}
\bibinfo{author}{Shimizu, K.}, \bibinfo{author}{Otsuka, R.},
  \bibinfo{author}{Hara, M.}, \bibinfo{author}{Minamitani, E.} \&
  \bibinfo{author}{Watanabe, S.}
\newblock \bibinfo{title}{Prediction of born effective charges using neural
  network to study ion migration under electric fields: applications to
  crystalline and amorphous li$_3$po$_4$} (\bibinfo{year}{2023}).
\newblock \eprint{arXiv:2305.19546}.

\bibitem{Born_ferroelectric}
\bibinfo{author}{Roy, A.}, \bibinfo{author}{Prasad, R.},
  \bibinfo{author}{Auluck, S.} \& \bibinfo{author}{Garg, A.}
\newblock \bibinfo{title}{First-principles calculations of born effective
  charges and spontaneous polarization of ferroelectric bismuth titanate}.
\newblock \emph{\bibinfo{journal}{Journal of Physics: Condensed Matter}}
  \textbf{\bibinfo{volume}{22}}, \bibinfo{pages}{165902}
  (\bibinfo{year}{2010}).
\newblock \urlprefix\url{http://dx.doi.org/10.1088/0953-8984/22/16/165902}.

\bibitem{Batatia2022Design}
\bibinfo{author}{Batatia, I.} \emph{et~al.}
\newblock \bibinfo{title}{The design space of e(3)-equivariant atom-centered
  interatomic potentials} (\bibinfo{year}{2022}).
\newblock \eprint{2205.06643}.

\bibitem{jax2018github}
\bibinfo{author}{Bradbury, J.} \emph{et~al.}
\newblock \bibinfo{title}{{JAX}: composable transformations of
  {P}ython+{N}um{P}y programs} (\bibinfo{year}{2018}).
\newblock \urlprefix\url{http://github.com/google/jax}.

\bibitem{e3nn_jax}
\bibinfo{author}{Geiger, M.}, \bibinfo{author}{Daigavane, A.},
  \bibinfo{author}{Kim, S.} \& \bibinfo{author}{Jamali, K.}
\newblock \bibinfo{title}{e3nn/e3nn-jax repository on github}
  (\bibinfo{year}{2023}).

\bibitem{jraph2020github}
\bibinfo{author}{Godwin*, J.} \emph{et~al.}
\newblock \bibinfo{title}{{J}raph: {A} library for graph neural networks in
  jax.} (\bibinfo{year}{2020}).
\newblock \urlprefix\url{http://github.com/deepmind/jraph}.

\bibitem{haiku2020github}
\bibinfo{author}{Hennigan, T.}, \bibinfo{author}{Cai, T.},
  \bibinfo{author}{Norman, T.}, \bibinfo{author}{Martens, L.} \&
  \bibinfo{author}{Babuschkin, I.}
\newblock \bibinfo{title}{{H}aiku: {S}onnet for {JAX}} (\bibinfo{year}{2020}).
\newblock \urlprefix\url{http://github.com/deepmind/dm-haiku}.

\bibitem{phonon_DDB_data}
\bibinfo{author}{Petretto, G.} \emph{et~al.}
\newblock \bibinfo{title}{High-throughput density-functional perturbation
  theory phonons for inorganic materials}.
\newblock \emph{\bibinfo{journal}{Scientific Data}}
  \textbf{\bibinfo{volume}{5}}, \bibinfo{pages}{180065} (\bibinfo{year}{2018}).

\bibitem{Si_INS}
\bibinfo{author}{Kim, D.~S.} \emph{et~al.}
\newblock \bibinfo{title}{Nuclear quantum effect with pure anharmonicity and
  the anomalous thermal expansion of silicon}.
\newblock \emph{\bibinfo{journal}{Proceedings of the National Academy of
  Sciences}} \textbf{\bibinfo{volume}{115}}, \bibinfo{pages}{1992–1997}
  (\bibinfo{year}{2018}).
\newblock \urlprefix\url{http://dx.doi.org/10.1073/pnas.1707745115}.

\bibitem{pbe}
\bibinfo{author}{Perdew, J.~P.}, \bibinfo{author}{Burke, K.} \&
  \bibinfo{author}{Ernzerhof, M.}
\newblock \bibinfo{title}{Generalized gradient approximation made simple}.
\newblock \emph{\bibinfo{journal}{Phys. Rev. Lett.}}
  \textbf{\bibinfo{volume}{77}}, \bibinfo{pages}{3865} (\bibinfo{year}{1996}).

\bibitem{VASP-PBEsol}
\bibinfo{author}{Perdew, J.~P.} \emph{et~al.}
\newblock \bibinfo{title}{Restoring the density-gradient expansion for exchange
  in solids and surfaces}.
\newblock \emph{\bibinfo{journal}{Phys. Rev. Lett.}}
  \textbf{\bibinfo{volume}{100}}, \bibinfo{pages}{136406}
  (\bibinfo{year}{2008}).
\newblock
  \urlprefix\url{https://link.aps.org/doi/10.1103/PhysRevLett.100.136406}.

\bibitem{vasp1}
\bibinfo{author}{Kresse, G.} \& \bibinfo{author}{Furthm\"uller, J.}
\newblock \bibinfo{title}{Efficient iterative schemes for ab initio
  total-energy calculations using a plane-wave basis set}.
\newblock \emph{\bibinfo{journal}{Phys. Rev. B}} \textbf{\bibinfo{volume}{54}},
  \bibinfo{pages}{11169} (\bibinfo{year}{1996}).

\bibitem{vasp2}
\bibinfo{author}{Kresse, G.} \& \bibinfo{author}{Furthm\"uller, J.}
\newblock \bibinfo{title}{Efficiency of ab-initio total energy calculations for
  metals and semiconductors using a plane-wave basis set}.
\newblock \emph{\bibinfo{journal}{Comput. Mater. Sci.}}
  \textbf{\bibinfo{volume}{6}}, \bibinfo{pages}{15} (\bibinfo{year}{1996}).

\bibitem{flat_phonon_ferroelectric}
\bibinfo{author}{Lee, H.-J.} \emph{et~al.}
\newblock \bibinfo{title}{Scale-free ferroelectricity induced by flat phonon
  bands in hfo2}.
\newblock \emph{\bibinfo{journal}{Science}} \textbf{\bibinfo{volume}{369}},
  \bibinfo{pages}{1343--1347} (\bibinfo{year}{2020}).
\newblock \urlprefix\url{https://doi.org/10.1126/science.aba0067}.

\bibitem{INS_dataset}
\bibinfo{author}{Cheng, Y.}, \bibinfo{author}{Stone, M.~B.} \&
  \bibinfo{author}{Ramirez-Cuesta, A.~J.}
\newblock \bibinfo{title}{A database of synthetic inelastic neutron scattering
  spectra from molecules and crystals}.
\newblock \emph{\bibinfo{journal}{Scientific Data}}
  \textbf{\bibinfo{volume}{10}}, \bibinfo{pages}{54} (\bibinfo{year}{2023}).

\bibitem{posym}
\bibinfo{author}{Carreras, A.}
\newblock \bibinfo{title}{Posym: A python library to analyze the symmetry of
  theoretical chemistry objects} (\bibinfo{year}{2022}).

\bibitem{Raman_tensor}
\bibinfo{author}{Bagheri, M.} \& \bibinfo{author}{Komsa, H.-P.}
\newblock \bibinfo{title}{High-throughput computation of raman spectra from
  first principles}.
\newblock \emph{\bibinfo{journal}{Scientific Data}}
  \textbf{\bibinfo{volume}{10}} (\bibinfo{year}{2023}).
\newblock \urlprefix\url{http://dx.doi.org/10.1038/s41597-023-01988-5}.

\bibitem{Force_not_enough}
\bibinfo{author}{Fu, X.} \emph{et~al.}
\newblock \bibinfo{title}{Forces are not enough: Benchmark and critical
  evaluation for machine learning force fields with molecular simulations}.
\newblock \emph{\bibinfo{journal}{TMLR}}  (\bibinfo{year}{2022}).
\newblock \eprint{arXiv:2210.07237}.

\bibitem{CCSD1}
\bibinfo{author}{Evangelista, F.~A.}, \bibinfo{author}{Prochnow, E.},
  \bibinfo{author}{Gauss, J.} \& \bibinfo{author}{Schaefer, H.~F.}
\newblock \bibinfo{title}{Perturbative triples corrections in state-specific
  multireference coupled cluster theory}.
\newblock \emph{\bibinfo{journal}{The Journal of Chemical Physics}}
  \textbf{\bibinfo{volume}{132}} (\bibinfo{year}{2010}).
\newblock \urlprefix\url{http://dx.doi.org/10.1063/1.3305335}.

\bibitem{CCSD2}
\bibinfo{author}{Evangelista, F.~A.}, \bibinfo{author}{Simmonett, A.~C.},
  \bibinfo{author}{Allen, W.~D.}, \bibinfo{author}{Schaefer, H.~F.} \&
  \bibinfo{author}{Gauss, J.}
\newblock \bibinfo{title}{Triple excitations in state-specific multireference
  coupled cluster theory: Application of mk-mrccsdt and mk-mrccsdt-n methods to
  model systems}.
\newblock \emph{\bibinfo{journal}{The Journal of Chemical Physics}}
  \textbf{\bibinfo{volume}{128}} (\bibinfo{year}{2008}).
\newblock \urlprefix\url{http://dx.doi.org/10.1063/1.2834927}.

\bibitem{Config_interact}
\bibinfo{author}{David~Sherrill, C.} \& \bibinfo{author}{Schaefer, H.~F.}
\newblock \emph{\bibinfo{title}{The Configuration Interaction Method: Advances
  in Highly Correlated Approaches}}, \bibinfo{pages}{143–269}
  (\bibinfo{publisher}{Elsevier}, \bibinfo{year}{1999}).
\newblock \urlprefix\url{http://dx.doi.org/10.1016/S0065-3276(08)60532-8}.

\bibitem{spectra_opt_transport}
\bibinfo{author}{Seifert, N.~A.}, \bibinfo{author}{Prozument, K.} \&
  \bibinfo{author}{Davis, M.~J.}
\newblock \bibinfo{title}{Computational optimal transport for molecular
  spectra: The fully continuous case}.
\newblock \emph{\bibinfo{journal}{The Journal of Chemical Physics}}
  \textbf{\bibinfo{volume}{159}} (\bibinfo{year}{2023}).
\newblock \urlprefix\url{http://dx.doi.org/10.1063/5.0166469}.

\bibitem{INS_ML}
\bibinfo{author}{Cheng, Y.} \emph{et~al.}
\newblock \bibinfo{title}{Direct prediction of inelastic neutron scattering
  spectra from the crystal structure}.
\newblock \emph{\bibinfo{journal}{Mach. Learn. Sci. Technol.}}
  \textbf{\bibinfo{volume}{4}}, \bibinfo{pages}{015010} (\bibinfo{year}{2023}).

\bibitem{magnetoelectric}
\bibinfo{author}{Eerenstein, W.}, \bibinfo{author}{Mathur, N.~D.} \&
  \bibinfo{author}{Scott, J.~F.}
\newblock \bibinfo{title}{Multiferroic and magnetoelectric materials}.
\newblock \emph{\bibinfo{journal}{Nature}} \textbf{\bibinfo{volume}{442}},
  \bibinfo{pages}{759–765} (\bibinfo{year}{2006}).
\newblock \urlprefix\url{http://dx.doi.org/10.1038/nature05023}.

\bibitem{Universal_MLpot}
\bibinfo{author}{Zhang, Y.} \& \bibinfo{author}{Jiang, B.}
\newblock \bibinfo{title}{Universal machine learning for the response of
  atomistic systems to external fields}.
\newblock \emph{\bibinfo{journal}{Nature Communications}}
  \textbf{\bibinfo{volume}{14}}, \bibinfo{pages}{6424} (\bibinfo{year}{2023}).

\bibitem{LLM_emergent}
\bibinfo{author}{Wei, J.} \emph{et~al.}
\newblock \bibinfo{title}{Emergent abilities of large language models}.
\newblock \emph{\bibinfo{journal}{Transactions on Machine Learning Research}}
  (\bibinfo{year}{2022}).
\newblock \urlprefix\url{https://openreview.net/forum?id=yzkSU5zdwD}.
\newblock \bibinfo{note}{Survey Certification}.

\bibitem{Phoebe}
\bibinfo{author}{Cepellotti, A.}, \bibinfo{author}{Coulter, J.},
  \bibinfo{author}{Johansson, A.}, \bibinfo{author}{Fedorova, N.~S.} \&
  \bibinfo{author}{Kozinsky, B.}
\newblock \bibinfo{title}{Phoebe: a high-performance framework for solving
  phonon and electron boltzmann transport equations}.
\newblock \emph{\bibinfo{journal}{Journal of Physics: Materials}}
  \textbf{\bibinfo{volume}{5}}, \bibinfo{pages}{035003} (\bibinfo{year}{2022}).

\bibitem{magnon_phonon_coupling}
\bibinfo{author}{Berk, C.} \emph{et~al.}
\newblock \bibinfo{title}{Strongly coupled magnon--phonon dynamics in a single
  nanomagnet}.
\newblock \emph{\bibinfo{journal}{Nature Communications}}
  \textbf{\bibinfo{volume}{10}}, \bibinfo{pages}{2652} (\bibinfo{year}{2019}).

\bibitem{reuther2018interactive}
\bibinfo{author}{Reuther, A.} \emph{et~al.}
\newblock \bibinfo{title}{Interactive supercomputing on 40,000 cores for
  machine learning and data analysis}.
\newblock In \emph{\bibinfo{booktitle}{2018 IEEE High Performance extreme
  Computing Conference (HPEC)}}, \bibinfo{pages}{1--6}
  (\bibinfo{organization}{IEEE}, \bibinfo{year}{2018}).

\end{thebibliography}

\onecolumngrid
\newpage

\end{document}